\documentclass[aps, showpacs, showkeys, nofootinbib, floatfix]{revtex4}
\usepackage{amssymb}
\usepackage{amsmath}
\usepackage{graphicx}
\usepackage{color}

\begin{document}

\title{Coupled dark energy with perturbed Hubble expansion rate}

\author{Weiqiang Yang$^a$}
\author{Lixin Xu$^{a,}$$^b$\footnote{lxxu@dlut.edu.cn}}

\affiliation{$^a$Institute of Theoretical Physics, School of Physics and Optoelectronic Technology, Dalian University of Technology, Dalian, 116024, P. R. China \\
$^b$State Key Laboratory of Theoretical Physics, Institute of Theoretical Physics, Chinese Academy of Sciences, Beijing, 100190, P. R. China
}

\begin{abstract}
The coupling between dark sectors provides a possible approach to mitigate the coincidence problem of cosmological standard model. In this paper, dark energy is treated as a fluid with a constant equation of state, whose coupling with dark matter is proportional the Hubble parameter and energy density of dark energy, that is, $\bar{Q}=3\xi_x\bar{H}\bar{\rho}_x$. Particularly, we consider the Hubble expansion rate to be perturbed in the perturbation evolutions of dark sectors. Using jointing data sets which include cosmic microwave background radiation, baryon acoustic oscillation, type Ia supernovae, and redshift-space distortions, we perform a full Monte Carlo Markov Chain likelihood analysis for the coupled model. The results show that the mean value with errors of interaction rate is: $\xi_x=0.00305_{-0.00305-0.00305-0.00305}^{+0.000645+0.00511+0.00854}$ for $Q^{\mu}_A\parallel u^{\mu}_c$; $\xi_x=0.00317_{-0.00317-0.00317-0.00317}^{+0.000628+0.00547+0.00929}$ for $Q^{\mu}_A\parallel u^{\mu}_x$, which means that the recently cosmic observations favored small interaction rate which is up to the order of $10^{-3}$. Moreover, in contrast to the coupled model with unperturbed expansion rate, we find perturbed Hubble expansion rate could bring about negligible impact on the model parameter space.
\end{abstract}

\pacs{98.80.-k, 98.80.Es}
\maketitle

\section{Introduction}

The analysis on the cosmological standard model from the Planck data \cite{ref:Planck2013,ref:Planck2013-CMB,ref:Planck2013-params} tells us that dark energy occupies about $68.3\%$ of the Universe, dark matter accounts for $26.8\%$, and baryonic matter occupies $4.9\%$. Although the $\Lambda$CDM model, which is consisted of the cosmological constant and cold dark matter, is in good agreement with the cosmic observational data, it meets with the coincidence problem \cite{ref:Zlatev1999,ref:Chimento2003,ref:Huey2006}.
An effective method to alleviate this issue is to consider the coupling between dark matter and dark energy. The coupling of dark sectors would influence the background evolution of the Universe and affect the growth history of the cosmic structure. Up to now, it is different to identify the coupled form from the fundamental theory. Thus, the coupled dark energy models are mostly on the phenomenological consideration.
We roughly class these interaction forms into two types. The first type of coupled model is only related to the energy densities of dark fluids \cite{ref:Potter2011,ref:Aviles2011,
ref:Caldera-Cabral2009,ref:Caldera-Cabral2009-2,ref:Boehmer2010,ref:Boehmer2008,ref:Song2009,ref:Koyama2009,
ref:Majerotto2010,ref:Valiviita2010,ref:Valiviita2008,ref:Jackson2009,ref:Clemson2012}. The second type of coupled model is proportional to the Hubble expansion rate and energy densities of dark sectors. \cite{ref:Chongchitnan2009,ref:Gavela2009,ref:Gavela2010,ref:Salvatelli2013,ref:Yang2014-uc,ref:Yang2014-ux,
ref:Quartin2008,ref:Honorez2010,ref:Costa2013,ref:Bernardis2011,ref:He2011,ref:He2010,
ref:He2008,ref:Abdalla2009,ref:Sadjadi2010,ref:Olivares2008,ref:Olivares2006,
ref:Olivares2005,ref:Sun2013,ref:Sadjadi2006,ref:Sadeghi2013,ref:Zhang2013,ref:Koivisto2005,ref:Simpson2011,
ref:Bertolami2007,ref:Avelino2012,ref:Quercellini2008,ref:Mohammadi2012,ref:Sharif2012,ref:Wu2007,
ref:Barrow2006,ref:Zimdahl2001,
ref:Koshelev2009,ref:Liyh2013,ref:Chimento2013,ref:Martin2014}.

The coupling between dark sectors could significantly affect the growth history of cosmic structure, one can see Refs. \cite{ref:Clemson2012,ref:Caldera-Cabral2009,ref:Song2009,ref:Koyama2009,ref:Honorez2010,ref:Koshelev2009}. In the test of galaxy clustering, the redshifts need to be translated to distances, so the measured clustering would be highly anisotropic. An important source of this anisotropy are redshift-space distortions (RSD) \cite{ref:RSD-Kaiser1987}. RSD arise because peculiar velocities contribute to observed galaxies redshifts, a spherical overdensity appears distorted by peculiar velocities when observed in the redshift space. On linear scales, the overdenstiy appears squashed along the line of sight. For a detailed review of RSD, one can see Ref. \cite{ref:RSD-Hamilton1998}. RSD allow measurements of the amplitude of fluctuations in the velocity field, in linear theory, a model-dependent measurement of $f\sigma_8(z)$ has been suggested in Ref. \cite{ref:fsigma8-DE-Song2009}, where $\sigma_8(z)$ is the overall normalisation of the matter density fluctuations. Many RSD measurements have been summarized from a variety of galaxy surveys in Table \ref{tab:fsigma8data}, including the 2dFGRS \cite{ref:fsigma81-Percival2004}, the WiggleZ \cite{ref:fsigma82-Blake2011}, the SDSS LRG \cite{ref:fsigma83-Samushia2012}, the BOSS CMASS \cite{ref:fsigma84-Reid2012}, the 6dFGRS \cite{ref:fsigma85-Beutler2012}, and the VIPERS \cite{ref:fsigma86-Torre2013} surveys. These measurements have in turn been used to set constraints on the cosmological growth rate. Moreover, a lower growth rate from RSD than expected from Planck data was also pointed out in Ref. \cite{ref:fsigma87-Macaulay2013}.

\begin{table}
\begin{center}
\begin{tabular}{ccc}
\hline\hline z & $f\sigma_8(z)$ & survey and references \\ \hline
$0.067$ & $0.42\pm0.06$ & $6dFGRS~(2012)$ \cite{ref:fsigma85-Beutler2012}\\
$0.17$ & $0.51\pm0.06$ & $2dFGRS~(2004)$ \cite{ref:fsigma81-Percival2004}\\
$0.22$ & $0.42\pm0.07$ & $WiggleZ~(2011)$ \cite{ref:fsigma82-Blake2011}\\
$0.25$ & $0.39\pm0.05$ & $SDSS~LRG~(2011)$ \cite{ref:fsigma83-Samushia2012}\\
$0.37$ & $0.43\pm0.04$ & $SDSS~LRG~(2011)$ \cite{ref:fsigma83-Samushia2012}\\
$0.41$ & $0.45\pm0.04$ & $WiggleZ~(2011)$ \cite{ref:fsigma82-Blake2011}\\
$0.57$ & $0.43\pm0.03$ & $BOSS~CMASS~(2012)$ \cite{ref:fsigma84-Reid2012}\\
$0.60$ & $0.43\pm0.04$ & $WiggleZ~(2011)$ \cite{ref:fsigma82-Blake2011}\\
$0.78$ & $0.38\pm0.04$ & $WiggleZ~(2011)$ \cite{ref:fsigma82-Blake2011}\\
$0.80$ & $0.47\pm0.08$ & $VIPERS~(2013)$ \cite{ref:fsigma86-Torre2013}\\
\hline\hline
\end{tabular}
\caption{The data points of $f\sigma_8(z)$ measured from RSD with the survey references.}
\label{tab:fsigma8data}
\end{center}
\end{table}

The RSD-derived $f\sigma_8(z)$ tests provide a very powerful and robust test of both the nature of dark energy and modified gravity, and could be used to constrain basic cosmological parameters. What's more, $f\sigma_8(z)$ measurements could significantly enhances the constraints on the model parameter space compared to the case where only geometric tests are used. The jointing constraints on several models have been tested in Refs. \cite{ref:fsigma8-HDE-Xu2013,ref:fsigma8andPlanck-MG-Xu2013,ref:Xu2013-DGP,ref:Yang2013-wdm,ref:Yang2013-cass,
ref:fsigma8total-Samushia2013}. For the coupled dark energy model, the geometry measurements mildly favor the coupling between dark sectors \cite{ref:Clemson2012,ref:Salvatelli2013,ref:Costa2013,ref:Chimento2013}, at the same time, the measurement about the growth rate of dark matter perturbations could rule out large interaction rate. Under the inspiration of this idea, Yang and Xu combined the geometric tests with the RSD measurement to constrain the dark energy model when the momentum transfer was vanished in the rest frame of dark matter or dark energy \cite{ref:Yang2014-uc,ref:Yang2014-ux}, the jointing data sets was able to estimate the parameter space to high precision and evidently tighten the constraints, they found cosmic observations favored small interaction rate, however, only the background expansion rate was considered. In this paper, we will extend this work to investigate the coupled dark energy model with perturbed Hubble expansion rate, which is inspired by Ref. \cite{ref:Gavela2010}. In this case, $H$ denotes the total expansion rate (background plus perturbations), $H=\bar{H}+\delta H$.
If we consider that the energy transfer rate is proportional to the Hubble parameter and the energy density of dark energy, the background coupling term is $\bar{Q}=3\xi_x\bar{H}\bar{\rho}_x$ ($\xi_x$ is called as the interaction rate), and the perturbation part of the interaction is $\delta Q=3\xi_x\bar{H}\bar{\rho}_x(\delta H/\bar{H}+\delta_x)$. This coupled form with regards to energy density of dark energy is free of large scale instability \cite{ref:Valiviita2008,ref:Clemson2012,ref:Gavela2009}. Furthermore, in the light of the analysis in Refs. \cite{ref:Gavela2009,ref:Clemson2012}, the stability conditions of the perturbations are $\xi_x>0$ and $(1+w_x)>0$. As for the phantom case $w_x<-1$, together with $\xi_x<0$, which does not suffer from the instability, but we exclude it on account of the unphysical property \cite{ref:Caldwell2003}.

This paper is organized as follows. In Sec. II, we consider the coupled model between dynamical dark energy and cold dark matter, hereafter, we call it as the $\xi$wCDM model. With perturbed Hubble expansion rate, we deduce the perturbation equations of dark sectors in the rest frame of dark matter or dark energy. In Sec. III, we present the cosmological implications of the interaction rate. Then, we constrain the coupled dark energy model with jointing the RSD measurements and geometry tests. The last section is the conclusions of this paper.

\section{The background and perturbation equations of dark fluids}

In a flat Friedman-Robertson-Walker (FRW) universe, according to the phenomenological approach, the coupling could be introduced into the background conservation equations of dark matter and dark energy
\begin{eqnarray}
\bar{\rho}'_c+3\mathcal{H}\bar{\rho}_c=a\bar{Q}_c=-a\bar{Q},
\label{eq:rhoc}
\end{eqnarray}
\begin{eqnarray}
\bar{\rho}'_x+3\mathcal{H}(1+w_x)\bar{\rho}_x=a\bar{Q}_x=a\bar{Q},
\label{eq:rhox}
\end{eqnarray}
where the subscript $c$ and $x$ respectively stand for dark matter and dark energy, the prime denotes the derivative with respect to conformal time $\tau$, $a$ is the scale factor of the Universe, $\mathcal{H}=a'/a=a\bar{H}$ is the conformal Hubble parameter, $w_x=\bar{p}_x/\bar{\rho}_x$ is the equation of state (EoS) parameter of dark energy. $\bar{Q}>0$ presents that the direction of energy transfer is from dark matter to dark energy, which changes the dark matter and dark energy redshift dependence acting as an extra contribution to their effective EoS; $\bar{Q}<0$ means the opposite direction of the energy exchange.

In a general gauge, the perturbed FRW metric is \cite{ref:Majerotto2010,ref:Valiviita2008,ref:Clemson2012}
\begin{eqnarray}
ds^2=a^2(\tau)\{ -(1+2\phi)d\tau^2+2\partial_iBd\tau dx^i+[(1-2\psi)\delta_{ij}+2\partial_i\partial_jE]dx^idx^j \},
\label{eq:per-metric}
\end{eqnarray}
where $\phi$, B, $\psi$ and E are the gauge-dependent scalar perturbations quantities.

The general forms for density perturbations (continuity) and velocity perturbations (Euler) equations of A fluid \cite{ref:Majerotto2010,ref:Valiviita2008,ref:Clemson2012}
\begin{eqnarray}
\delta'_A+3\mathcal{H}(c^2_{sA}-w_A)\delta_A
+9\mathcal{H}^2(1+w_A)(c^2_{sA}-c^2_{aA})\frac{\theta_A}{k^2}
+(1+w_A)\theta_A-3(1+w_A)\psi'+(1+w_A)k^2(B-E')
\nonumber \\
=\frac{a}{\bar{\rho}_A}(-\bar{Q}_A\delta_A+\delta Q_A)
+\frac{a\bar{Q}_A}{\bar{\rho}_A}\left[\phi+3\mathcal{H}(c^2_{sA}-c^2_{aA})\frac{\theta_A}{k^2}\right],
\label{eq:general-deltaA}
\end{eqnarray}
\begin{eqnarray}
\theta'_A+\mathcal{H}(1-3c^2_{sA})\theta_A-\frac{c^2_{sA}}{(1+w_A)}k^2\delta_A
-k^2\phi
=\frac{a}{(1+w_A)\bar{\rho}_A}[(\bar{Q}_A\theta-k^2f_A)-(1+c^2_{sA})\bar{Q}_A\theta_A],
\label{eq:general-thetaA}
\end{eqnarray}
where $\delta_A=\delta\rho_A/\bar{\rho}_A$ is the density contrast of A fluid, $\theta_A=-k^2(v_A+B)$ is the volume expansion of A fluid in Fourier space \cite{ref:Valiviita2008,ref:Ma1995}, $\theta$ is the volume expansion of total fluid, $v_A$ is the peculiar velocity potential, $k$ is the wavenumber; $c^2_{aA}$ is the adiabatic sound speed whose definition is $c^2_{aA}=\bar{p}'_A/\bar{\rho}'_A=w_x+w'_x/(\bar{\rho}'_A/\bar{\rho}_A)$, and $c^2_{sA}$ is the physical sound speed in the rest frame, its definition is $c^2_{sA}=(\delta p_A/\delta\rho_A)_{restframe}$ \cite{ref:Valiviita2008,ref:Kodama1984,ref:Hu1998,ref:Gordon2004}. In order to avoid the unphysical instability, $c^2_{sA}$ should be taken as a non-negative parameter \cite{ref:Valiviita2008}. $f_A$ is the momentum transfer potential, which is usually assumed that $k^2f_A=\bar{Q}_A[\theta-b\theta_c-(1-b)\theta_x]$ in the rest frame of dark matter ($b=1$ for $Q^{\mu}_A\parallel u^{\mu}_c$) or dark energy ($b=0$ for $Q^{\mu}_A\parallel u^{\mu}_x$) \cite{ref:Yang2014-uc}, where the energy-momentum transfer four-vector $Q^{\mu}_A$ is relative to the four-velocity $u^{\mu}$, and it can be split as $Q^A_0=-a[\bar{Q}_A(1+\phi)+\delta Q_A]$, $Q^A_i=a\partial_i[\bar{Q}_A(v+B)+f_A]$ \cite{ref:Majerotto2010,ref:Valiviita2008,ref:Clemson2012}.

When the perturbed Hubble expansion rate is considered in the perturbation equations of dark sectors, $H$ denotes the total expansion rate (background plus perturbations), $H=\bar{H}+\delta H$. In order to satisfy the gauge invariance of the theory, the expansion rate is chosen to be associated to the the volume expansion of total fluid \cite{ref:Gavela2010}. Then, in the synchronous gauge, the scalar perturbations are characterized by the metric perturbations $h$ and $\eta$ \cite{ref:Ma1995}, and the gauge transformation relations are $\phi=B=0$, $\psi=\eta$, and $k^2E=-h/2-3\eta$. According to the analysis on the contribution from the expansion rate perturbation $\delta H/\bar{H}$ in Ref. \cite{ref:Gavela2010}, $\delta H/\bar{H}=(\theta+h'/2)/(3\mathcal{H})$. Moreover, in light of $(\rho+p)v=\sum (\rho_A+p_A)v_A$ \cite{ref:Valiviita2008,ref:Gavela2010}, we could obtain the continuity and Euler equations of dark sectors \cite{ref:Yang2014-uc}

\begin{eqnarray}
\delta'_x+(1+w_x)\left(\theta_x+\frac{h'}{2}\right)+3\mathcal{H}(c^2_{sx}-w_x)\delta_x
+9\mathcal{H}^2(c^2_{sx}-w_x)(1+w_x)\frac{\theta_x}{k^2}
\nonumber \\
=9\mathcal{H}^2(c^2_{sx}-w_x)\xi_x\frac{\theta_x}{k^2}
+\xi_x\left(\theta+\frac{h'}{2}\right),
\label{eq:deltax-prime-b-dH-syn}
\end{eqnarray}
\begin{eqnarray}
\delta'_c+\theta_c+\frac{h'}{2}
=3\mathcal{H}\xi_x\frac{\rho_x}{\rho_c}(\delta_c-\delta_x)
-\xi_x\frac{\rho_x}{\rho_c}\left(\theta+\frac{h'}{2}\right),
\label{eq:deltac-prime-b-dH-syn}
\end{eqnarray}
\begin{eqnarray}
\theta'_x+\mathcal{H}(1-3c^2_{sx})\theta_x-\frac{c^2_{sx}}{1+w_x}k^2\delta_x
=\frac{3\mathcal{H}\xi_x}{1+w_x}
[b(\theta_c-\theta_x)-c^2_{sx}\theta_x],
\label{eq:thetax-prime-b-dH-syn}
\end{eqnarray}
\begin{eqnarray}
\theta'_c+\mathcal{H}\theta_c
=3\mathcal{H}\xi_x\frac{\rho_x}{\rho_c}(1-b)(\theta_c-\theta_x).
\label{eq:thetac-prime-b-dH-syn}
\end{eqnarray}
where perturbed expansion rate affects the perturbation equations by means of the last terms in Eqs. (\ref{eq:deltax-prime-b-dH-syn}) and (\ref{eq:deltac-prime-b-dH-syn}). Next, we would pay attention to the cosmological implications and constraint results of the interaction rate. Moreover, we try to find the difference between the coupled model with and without perturbed expansion rate.

\section{Cosmological implications and constraint results}

\subsection{Theoretical predictions of CMB temperature, matter power spectra, and $f\sigma_8(z)$}

\begin{figure}[tbh]
\includegraphics[width=8.8cm,height=6.8cm]{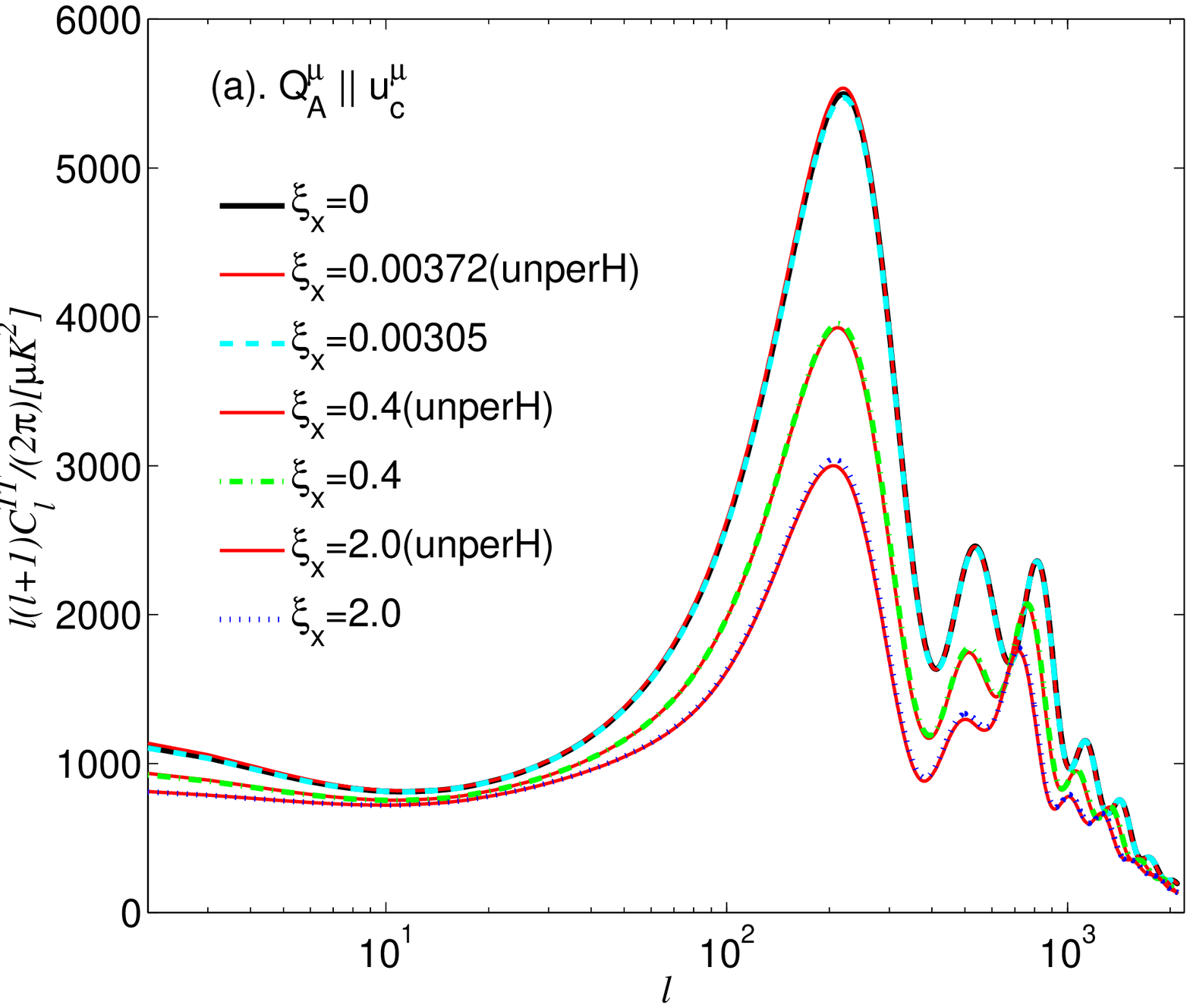}
\includegraphics[width=8.8cm,height=6.8cm]{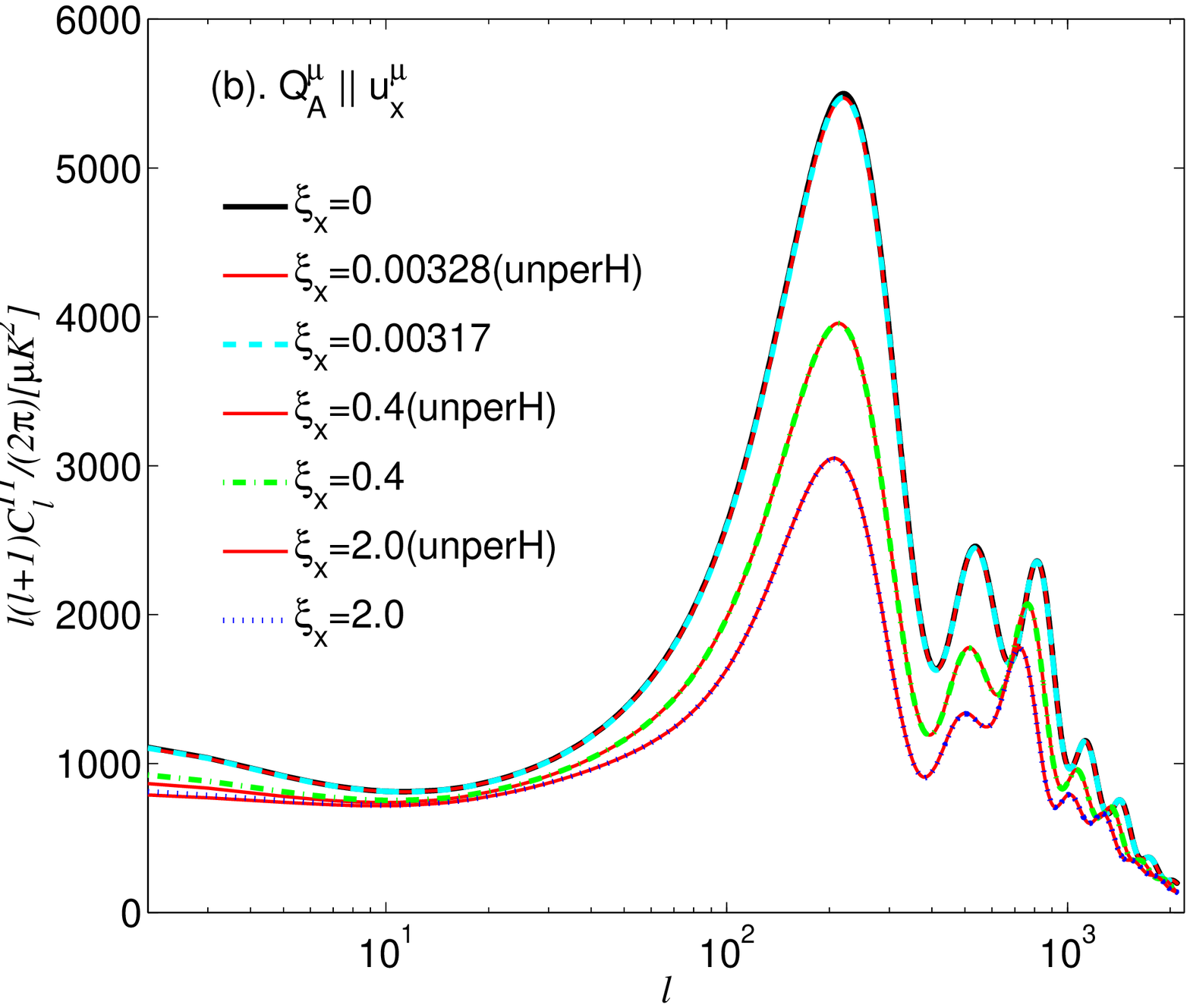}
 \caption{(a). The effects on CMB temperature power spectra for different values of interaction rate $\xi_x$ ($Q^{\mu}_A\parallel u^{\mu}_c$). The black solid, cyan thick dashed, green dotted-dashed, and blue dotted lines are for $\xi_x=0, 0.00305, 0.1$, and $0.2$, respectively; the other relevant parameters are fixed with the mean values as shown in the third column of Table \ref{tab:results-mean-ucdh}; the three thin red solid lines are the corresponding ones for the coupled model with unperturbed $H$ for $Q^{\mu}\parallel u^{\mu}_{(c)}$; (b). The corresponding evolutions for $Q^{\mu}_A\parallel u^{\mu}_x$, the relevant values of parameters is from Table \ref{tab:results-mean-uxdh}; the three thin red lines correspond to ones for the coupled model with unperturbed $H$ for $Q^{\mu}\parallel u^{\mu}_{(x)}$.}
\label{fig:CMBpower-dh}
\end{figure}

\begin{figure}[tbh]
\includegraphics[width=8.8cm,height=6.8cm]{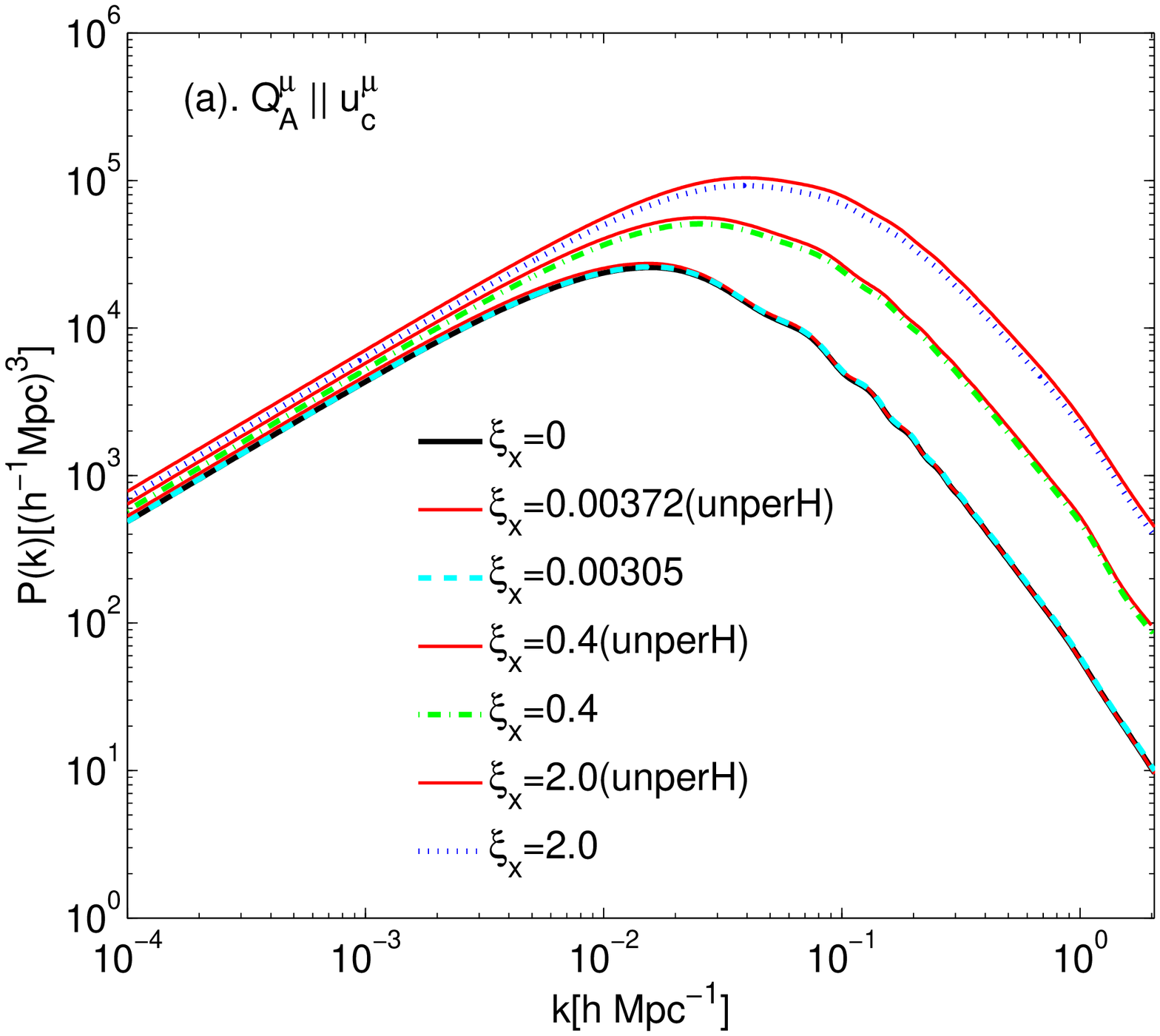}
\includegraphics[width=8.8cm,height=6.8cm]{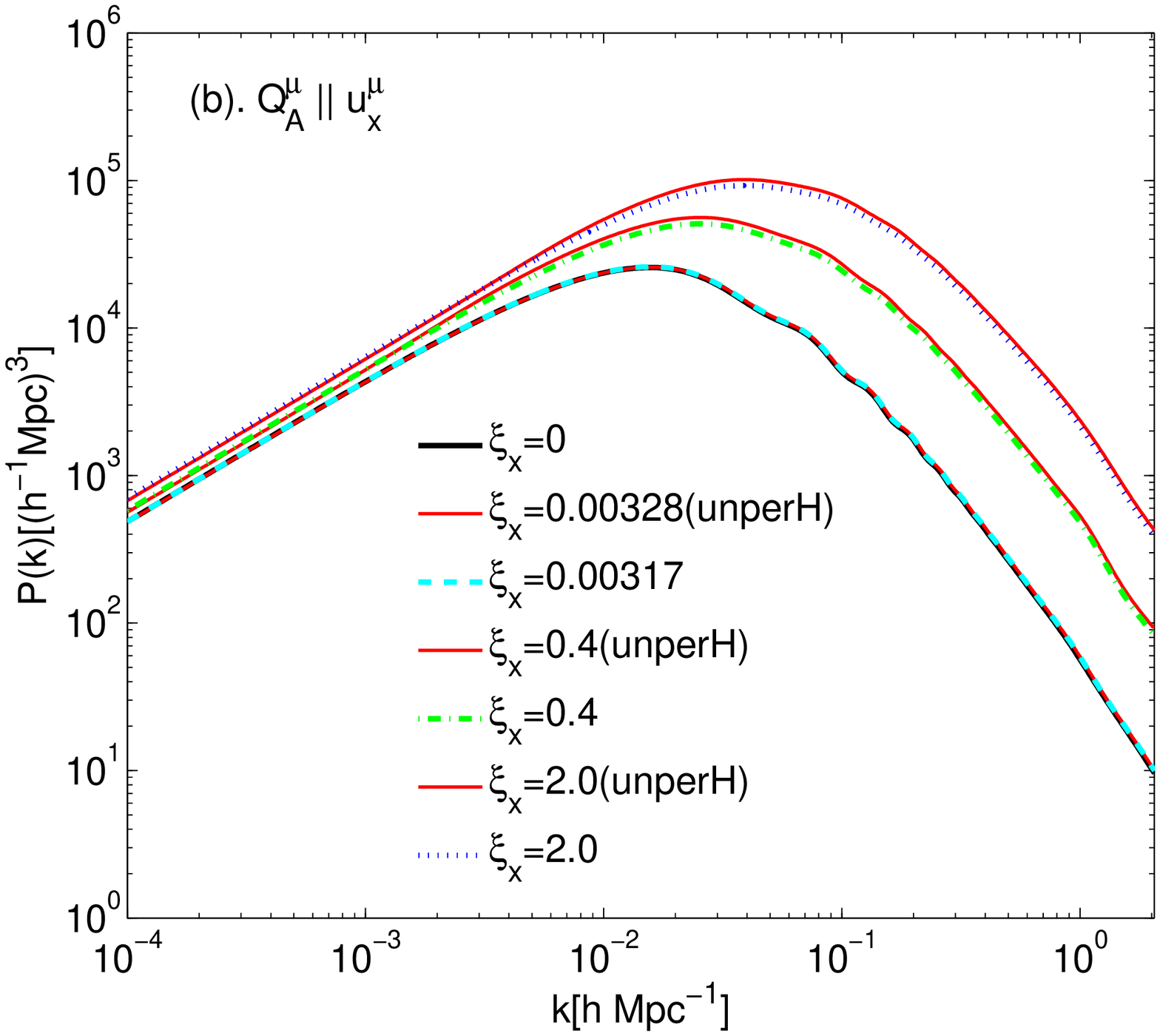}
 \caption{(a). The effects on matter power spectra for different values of interaction rate $\xi_x$ ($Q^{\mu}_A\parallel u^{\mu}_c$). The black solid, cyan thick dashed, green dotted-dashed, and blue dotted lines are for $\xi_x=0, 0.00305, 0.1$, and $0.2$, respectively; the other relevant parameters are fixed with the mean values as shown in the third column of Table \ref{tab:results-mean-ucdh}; the three thin red solid lines are the corresponding ones for the coupled model with unperturbed $H$ for $Q^{\mu}\parallel u^{\mu}_{(c)}$; (b). The corresponding evolutions for $Q^{\mu}_A\parallel u^{\mu}_x$, the relevant values of parameters is from Table \ref{tab:results-mean-uxdh}; the three thin red lines correspond to ones for the coupled model with unperturbed $H$ for $Q^{\mu}\parallel u^{\mu}_{(x)}$.}
\label{fig:Mpower-dh}
\end{figure}

In order to clearly show the cosmological implications of the interaction rate $\xi_x$, it is necessary to keep some model parameters fixed but only $\xi_x$ varied in the modified CAMB codes, so that we could find the cosmological effects of the coupling between dark energy and dark matter from the theoretical aspect. Firstly, we show the the influences on CMB temperature power spectra for varied interaction rate $\xi_x$ in Fig. \ref{fig:CMBpower-dh}. According to the background evolution equation of dark matter, with fixed density parameter $\Omega_c$ today, enlarging the values of positive $\xi_x$ would bring about greater corresponding $\Omega_{c}$ in the past, which could make the moment of matter-radiation equality earlier; therefore, the sound horizon is decreased. As a result, the first peak of CMB temperature spectra is depressed. As for the location shift of peaks, following the analysis about location of the CMB power spectra peaks in Ref. \cite{ref:Hu1995}, since the increasing $\xi_x$ is equivalent to enlarging $\Omega_m$, the peaks of power spectra would be shifted to smaller $l$. The similar case has occurred in Refs. \cite{ref:Clemson2012,ref:Yang2014-uc,ref:Yang2014-ux}. At large scales $l<100$, the integrated Sachs-Wolfe (ISW) effect is dominant, the changed parameter $\xi_x$ affects the CMB power spectra via ISW effect due to the evolution of gravitational potential. Here,
it is necessary to point out that the CMB power spectra own similar evolutions with ones of unperturbed $H$ coupled model (It is noted by gray lines in Fig. \ref{fig:CMBpower-dh}).
Then, we also plot the influences on the matter power spectrum $P(k)$ for the different values of interaction rate. With the increasment of $\xi_x$, the matter power spectra are enhanced at both lower and higher redshifts because of earlier moment of matter-radiation equality. With a large value of $\xi_x$,
there are little difference on the matter power spectra with ones of unperturbed $H$ coupled model (It is noted by gray lines in Fig. \ref{fig:Mpower-dh}), however it is not significant. Both the CMB temperature and matter power spectra are negligibly influenced by the interaction rate, even if perturbed expansion rate alters the perturbation equations by the last terms in Eqs. (\ref{eq:deltax-prime-b-dH-syn}) and (\ref{eq:deltac-prime-b-dH-syn}). Fortunately, the growth history of cosmic structure would be very sensitive for varied interaction rate.

\begin{figure}[tbh]
\includegraphics[width=8.8cm,height=6.8cm]{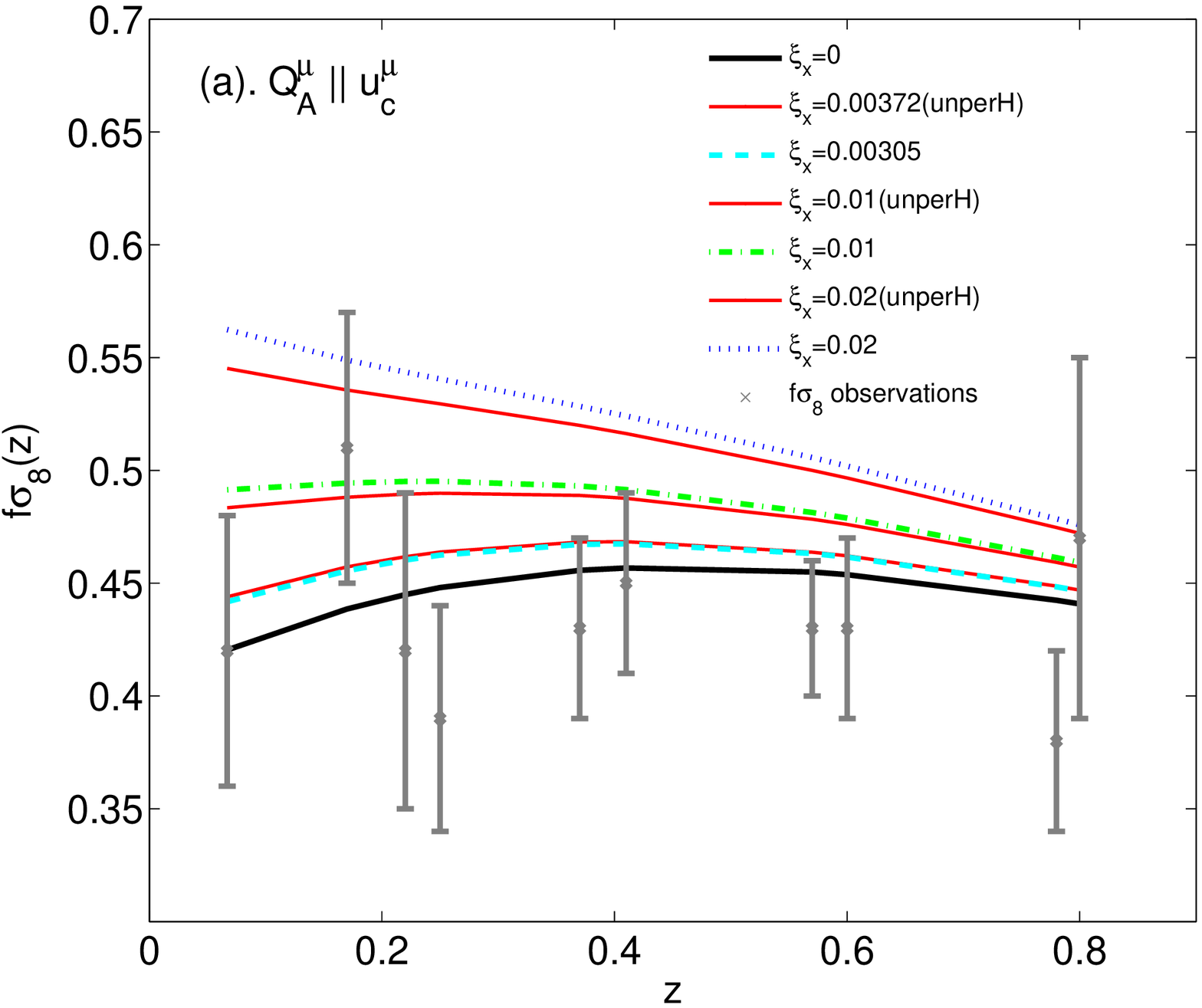}
\includegraphics[width=8.8cm,height=6.8cm]{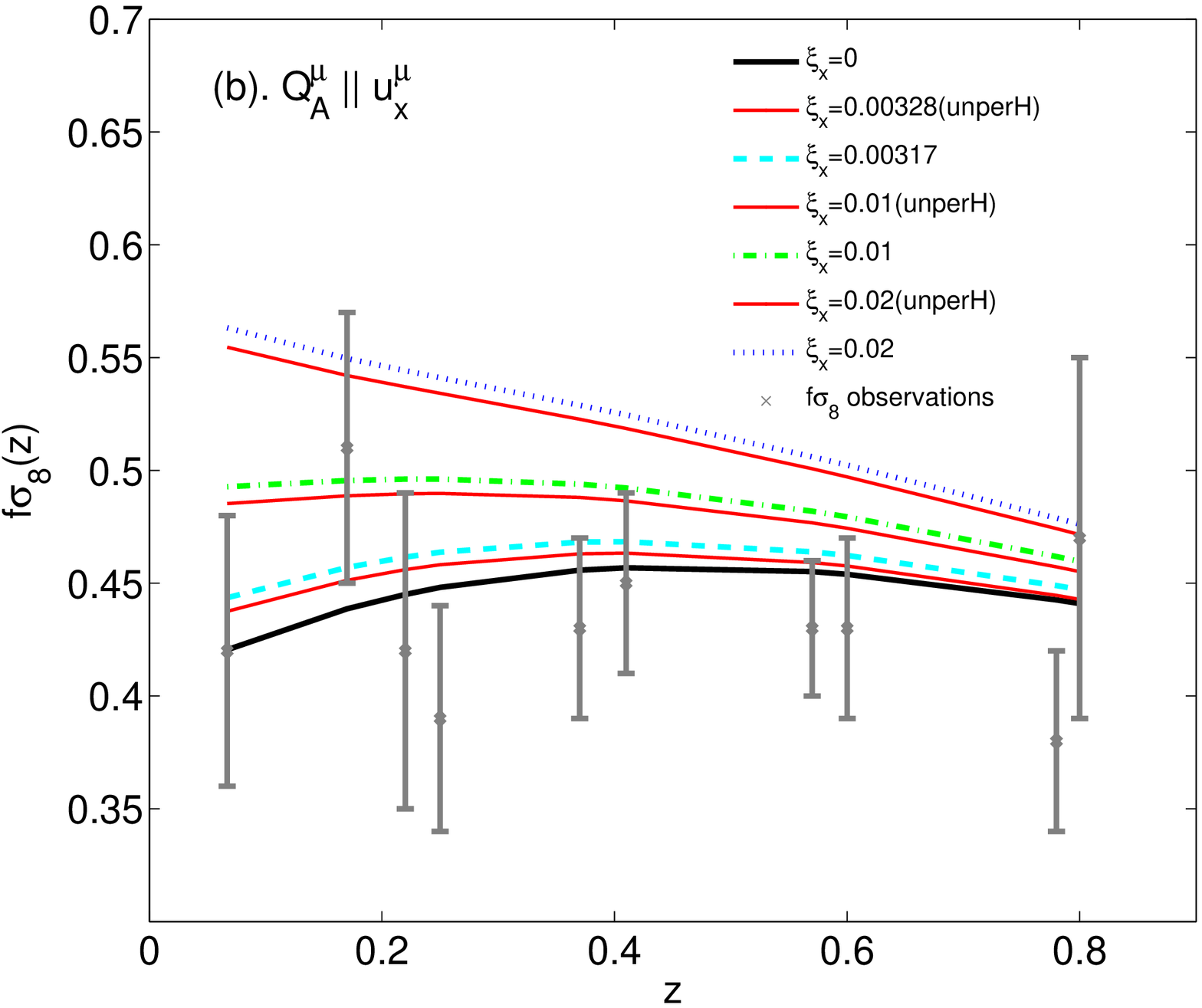}
 \caption{(a). The fitting evolutionary curves of $f\sigma_8(z)$ about the redshift $z$ for varied interaction rate $\xi_x$ ($Q^{\mu}_A\parallel u^{\mu}_c$). The black solid, cyan thick dashed, green dotted-dashed, and blue dotted lines are for $\xi_x=0, 0.00305, 0.01$, and $0.02$, respectively; the gray error bars denote the observations of $f\sigma_8(z)$ at different redshifts are listed in Table \ref{tab:fsigma8data}; the other relevant parameters are fixed with the mean values as shown in the third column of Table \ref{tab:results-mean-ucdh}; the three thin red solid lines are the corresponding ones for the coupled model with unperturbed $H$ for $Q^{\mu}\parallel u^{\mu}_{(c)}$; (b). The corresponding evolutions for $Q^{\mu}_A\parallel u^{\mu}_x$, the relevant values of parameters is from Table \ref{tab:results-mean-uxdh}; the three thin red lines correspond to ones for the coupled model with unperturbed $H$ for $Q^{\mu}\parallel u^{\mu}_{(x)}$.}
\label{fig:fsigma8-dh}
\end{figure}

In order to investigate the effects of interaction rate to $f\sigma_8(z)$, we fix the relevant mean values of our constraint results in Table \ref{tab:results-mean-ucdh} and Table \ref{tab:results-mean-uxdh}, but keep the model parameter $\xi_x$ varying in a range. In ten different redshifts (which could be used to compare with the observed $f\sigma_8$ data points), we derive the theoretical values of the growth rate from the new module in the modified CosmoMC package. When $\xi_x$ is fixed on a value, We fit the ten theoretical data points (z, $f\sigma_8(z)$) and plot the evolution curves of $f\sigma_8(z)$ in Fig. \ref{fig:fsigma8-dh}. At both lower and higher redshifts, the curves of $f\sigma_8(z)$ are enhanced with the increasing the values of $\xi_x$. Particularly, it is easy to see that the evolution curves of $f\sigma_8(z)$ are significantly distinguishing from ones of unperturbed $H$ coupled model (It is noted by gray lines in Fig. \ref{fig:fsigma8-dh}), even if the value of interaction rate is very small. We could qualitatively analyse why the evolution difference of $f\sigma_8(z)$ between these two models becomes large with reducing the redshift. For fixed $\xi_x$, at the higher redshift, the component of dark energy is subdominant, the last term of Eq. (\ref{eq:deltac-prime-b-dH-syn}) affecting the growth rate is trivial, which would slightly affect the evolution curves of $f\sigma_8$. Nonetheless, at the lower redshift, the dark energy gradually dominate the late Universe, the last term of Eq. (\ref{eq:deltac-prime-b-dH-syn}) would significantly influence the cosmic structure growth, which could bring about more obvious difference on the evolutions of $f\sigma_8(z)$ between the coupled model with or without perturbed $H$. To some extent, the RSD tests could break the possible degeneracy between perturbed $H$ coupled model and unperturbed $H$ coupled model.

\subsection{Data sets and constraint results}

For the $\xi$wCDM model, we consider the following eight-dimensional parameter space
\begin{eqnarray}
P\equiv\{\Omega_bh^2, \Omega_{c}h^2, \Theta_S, \tau, w_x, \xi_x, n_s, log[10^{10}A_S]\},
\label{eq:parameter_space}
\end{eqnarray}
the priors of the basic model parameters are shown in the second column of Table \ref{tab:results-mean-ucdh} or Table \ref{tab:results-mean-uxdh}. The pivot scale of the initial scalar power spectrum $k_{s0}=0.05Mpc^{-1}$ is adopted. Moreover, the priors of the cosmic age $10Gyr<t_0<20Gyr$ and Hubble constant $H_0=73.8\pm2.4 km s^{-1} Mpc^{-1}$ \cite{ref:Riess2011} are used. In order to avoid the unphysical sound speed, we assume $c^2_{sx}=1$ according to Refs. \cite{ref:Valiviita2008,ref:Majerotto2010,ref:Clemson2012}.

For our numerical calculations, the total likelihood $\chi^2$ can be constructed as
\begin{eqnarray}
\chi^2=\chi^2_{CMB}+\chi^2_{BAO}+\chi^2_{SNIa}+\chi^2_{RSD},
\label{eq:chi2}
\end{eqnarray}
where the four terms in right side of this equation, respectively, denote the contribution from CMB, BAO, SNIa, and RSD data sets. The used data sets for our Monte Carlo Markov Chain (MCMC) likelihood analysis are listed in Table \ref{tab:alldata}.

\begin{table}
\begin{center}
\begin{tabular}{ccc}
\hline\hline Data names & Data descriptions and references \\ \hline
CMB & $l\in[50,2500]$, high$-l$ temperature likelihood from Planck \cite{ref:Planck2013-params} \\
$...$ & $l\in[2,49]$, low$-l$ temperature likelihood from Planck \cite{ref:Planck2013-params} \\
$...$ & $l\in[2,32]$, low$-l$ polarization likelihood from WMAP9 \cite{ref:WMAP9} \\
BAO & $r_s/D_V(z=0.106)=0.336\pm0.015$ \cite{ref:BAO-1}\\
$...$ & $r_s/D_V(z=0.35)=0.1126\pm0.0022$ \cite{ref:BAO-2}\\
$...$ & $r_s/D_V(z=0.57)=0.0732\pm0.0012$ \cite{ref:BAO-3}\\
SNIa & SNLS3 data from SiFTO and SALT2 \cite{ref:SNLS3-1,ref:SNLS3-2}\\
RSD & ten $f\sigma_8(z)$ data points from Table \ref{tab:fsigma8data}\\
\hline\hline
\end{tabular}
\caption{The used data sets for our MCMC likelihood analysis on the coupled dark energy model, where $l$ is the multipole number of power spectra, and WMAP9 is the abbreviation of nine-year Wilkinson Microwave Anisotropy Probe.}
\label{tab:alldata}
\end{center}
\end{table}

\begin{figure}[!htbp]
\includegraphics[width=20cm,height=15cm]{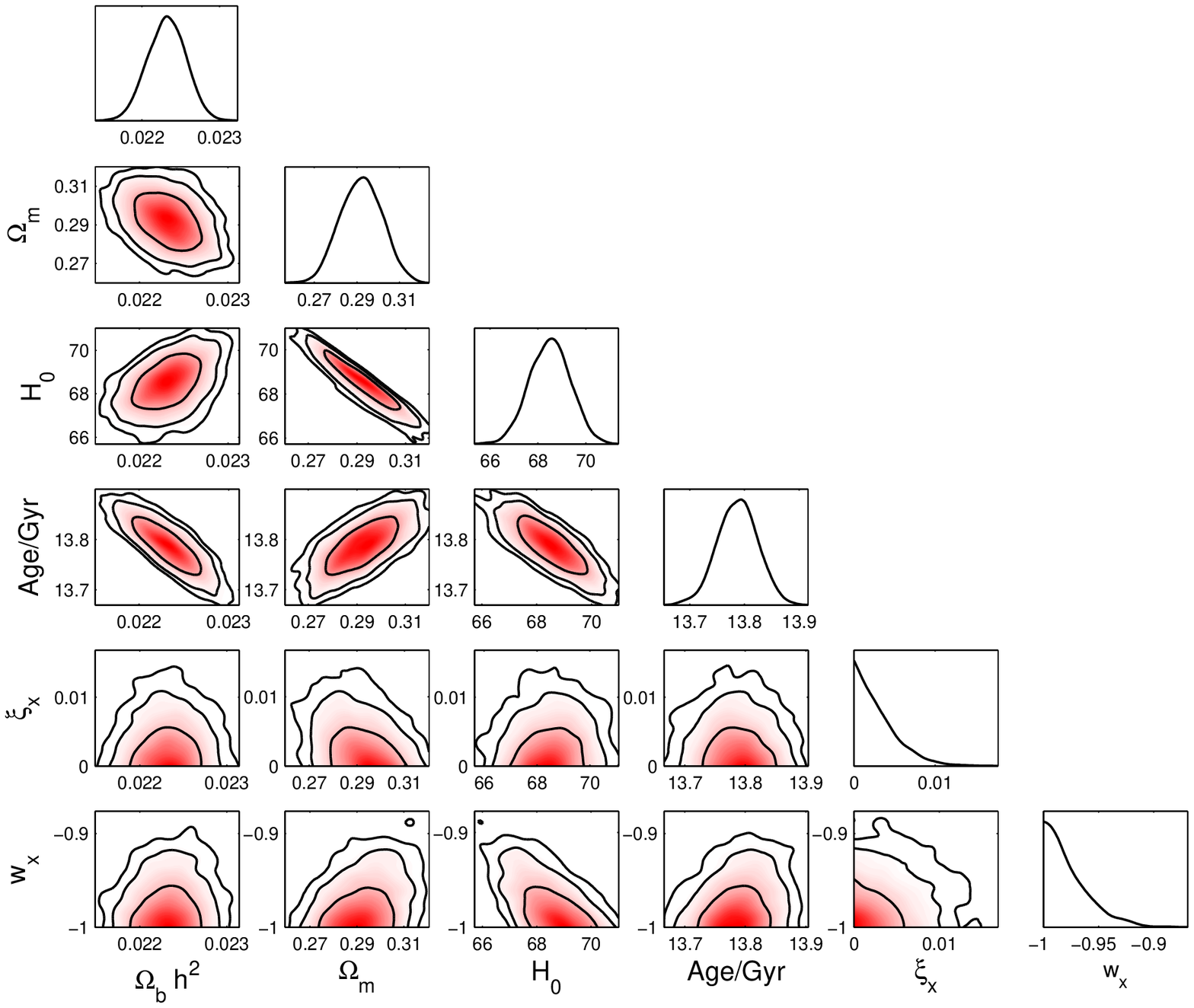}
  \caption{For the case of $Q^{\mu}_A\parallel u^{\mu}_c$, the 1D marginalized distributions on individual parameters and 2D contours with 68\% C.L., 95 \% C.L., and 99.7\% C.L. between each other using the combination of the observed data points from the CMB from Planck + WMAP9, BAO, SNIa, and RSD data sets.}
  \label{fig:contour-ucdh}
\end{figure}

\begin{figure}[!htbp]
\includegraphics[width=20cm,height=15cm]{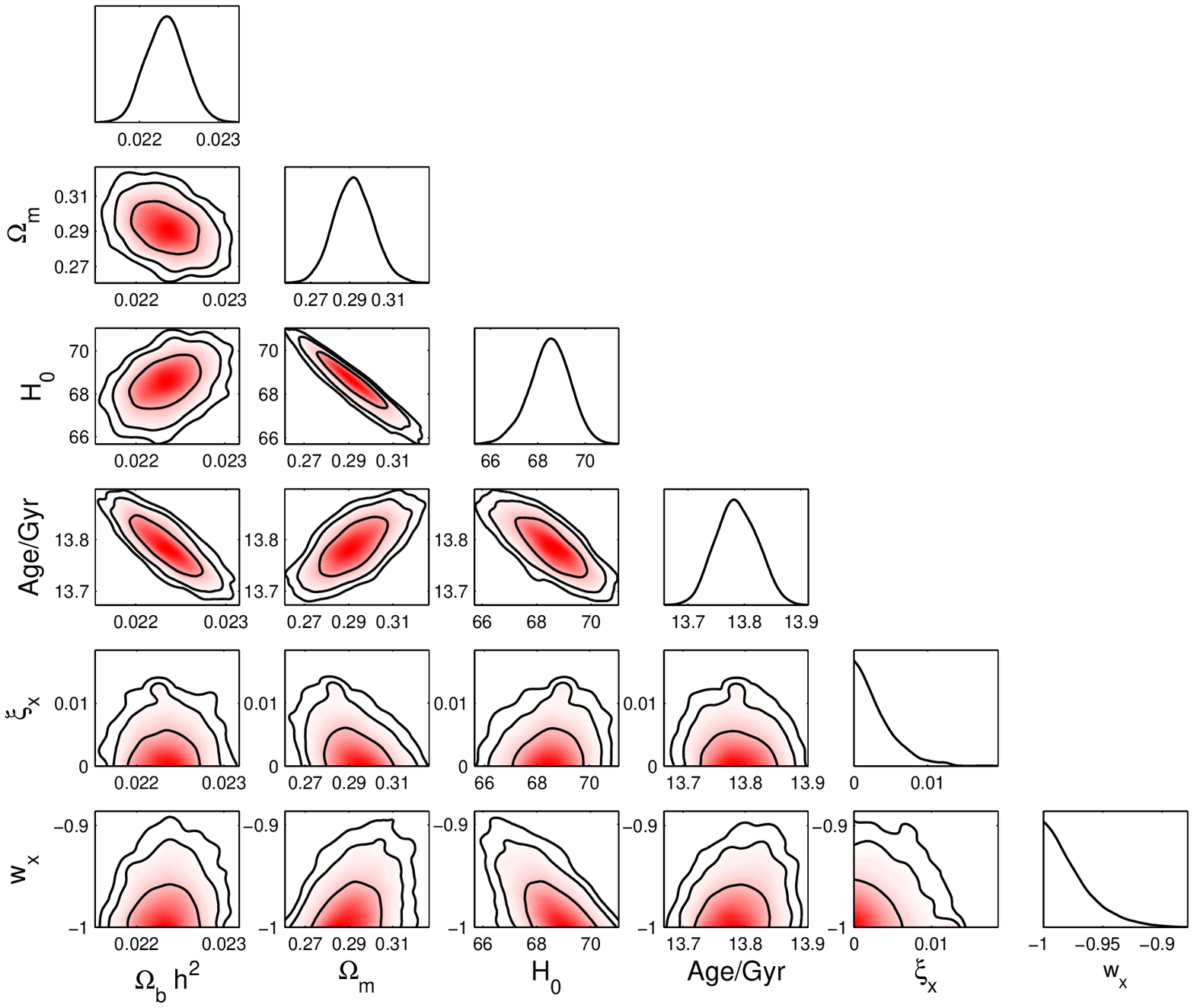}
  \caption{For the case of $Q^{\mu}_A\parallel u^{\mu}_x$, the 1D marginalized distributions on individual parameters and 2D contours with 68\% C.L., 95 \% C.L., and 99.7\% C.L. between each other using the combination of the observed data points from the CMB from Planck + WMAP9, BAO, SNIa, and RSD data sets.}
  \label{fig:contour-uxdh}
\end{figure}

After running eight chains in parallel, the mean values with errors and best-fit values, respectively, presented in the third and fourth columns of Table \ref{tab:results-mean-ucdh} and Table \ref{tab:results-mean-uxdh}. In Figs. \ref{fig:contour-ucdh} and \ref{fig:contour-uxdh}, We present the one-dimensional (1D) marginalized distributions of parameters and two-dimensional (2D) contours with $68\%$ confidence levels (C.L.), $95\%$ C.L., and $99.7\%$ C.L. In order to clearly see that the impact on the cosmological constraints of perturbed expansion rate, we also show the constraint results on $\xi$wCDM model with unperturbed $H$ coupled model in the fifth and sixth columns of Table \ref{tab:results-mean-ucdh} and Table \ref{tab:results-mean-uxdh}.


Firstly, we pay attention to the interaction rate in the coupled dark energy model. Using CMB from Planck + WMAP9, BAO, SNIa and RSD measurements, the results showed the interaction rate in 3$\sigma$ regions: $\xi_x=0.00305_{-0.00305-0.00305-0.00305}^{+0.000645+0.00511+0.00854}$ for $Q^{\mu}_A\parallel u^{\mu}_c$ and $\xi_x=0.00317_{-0.00317-0.00317-0.00317}^{+0.000628+0.00547+0.00929}$ for $Q^{\mu}_A\parallel u^{\mu}_x$. We find the recently cosmic observations indeed favor small interaction rate. Moreover, in 1$\sigma$ region, the $f\sigma_8(z)$ tests could rule out large interaction rate. Then, based on the same observed data sets (CMB from Planck + WMAP9, BAO, SNIa and RSD), in contrast to the constraint results of $\xi$wCDM model with unperturbed $H$, the new terms in the perturbation equations arising from the expansion rate perturbation have negligible quantitative impact on the constraints on cosmological parameters, even if perturbed expansion rate alters the perturbation equations by the last terms in Eqs. (\ref{eq:deltax-prime-b-dH-syn}) and (\ref{eq:deltac-prime-b-dH-syn}). We could draw the same conclusion from both the two coupled cases $Q^{\mu}_A\parallel u^{\mu}_c$ in Table \ref{tab:results-mean-ucdh} and $Q^{\mu}_A\parallel u^{\mu}_x$ in Table \ref{tab:results-mean-uxdh}. That is to say, the constraint results between the perturbed $H$ and unperturbed $H$ coupled model are compatible with each other. For these two coupled models with or without perturbed expansion rate, it would be also very hard to distinguish them.

\begingroup
\squeezetable
\begin{center}
\begin{table}
\begin{tabular}{cccccc}
\hline\hline Parameters & Priors & $\xi$wCDM with perturbed $H$ & Best fit & $\xi$wCDM with unperturbed $H$ & Best fit \\ \hline
$\Omega_bh^2$&[0.005,0.1]&
$0.0223_{-0.000245-0.000470-0.000616}^{+0.000241+0.000463+0.000602}$&$0.0222$&
$0.0223_{-0.000240-0.000490-0.000613}^{+0.000233+0.000490+0.000642}$&$0.0223$
\\
$\Omega_ch^2$&[0.01,0.99]&
$0.114_{-0.00169-0.00387-0.00552}^{+0.00214+0.00360+0.00417}$&$0.115$&
$0.114_{-0.00171-0.00405-0.00602}^{+0.00217+0.00385+0.00450}$&$0.115$
\\
$100\theta_{MC}$&[0.5,10]&
$1.0416_{-0.000551-0.00108-0.00139}^{+0.000569+0.00108+0.00140}$&$1.0412$&
$1.0416_{-0.000573-0.00113-0.00145}^{+0.000570+0.00111+0.00139}$&$1.0413$
\\
$\tau$&[0.01,0.8]&
$0.0870_{-0.0141-0.0238-0.0310}^{+0.0120+0.0261+0.0336}$&$0.0805$&
$0.0862_{-0.0122- 0.0226-0.0305}^{+0.0120+0.0239+0.0337}$&$0.0831$
\\
$\xi_x$&[0,1]&
$0.00305_{-0.00305-0.00305-0.00305}^{+0.000645+0.00511+0.00854}$&$0.000512$&
$0.00372_{-0.00372-0.00372- 0.00372}^{+0.000768+0.00655+0.0102}$&$0.00328$
\\
$w_x$&[-1,0]&
$-0.976_{-0.0237-0.0237-0.0237}^{+0.00503+0.0409+0.0612}$&$-0.989$&
$-0.975_{-0.0246-0.0246- 0.0246}^{+0.00581+0.0382+0.0601}$&$-0.995$
\\
$n_s$&[0.5,1.5]&
$0.977_{-0.00576-0.0109-0.0143}^{+0.00557+0.0111+0.0142}$&$0.977$&
$0.977_{-0.00550-0.0107- 0.0139}^{+0.00550+0.0109+0.0145}$&$0.975$
\\
${\rm{ln}}(10^{10}A_s)$&[2.4,4]&
$3.0812_{-0.0267-0.0459-0.0612}^{+0.0234+0.0494+0.0632}$&$3.0707$&
$3.0802_{-0.0232-0.0441-0.0603}^{+0.0229+0.0467+0.0642}$&$3.0784$
\\
\hline
$\Omega_x$&$-$&
$0.708_{-0.00987-0.0188-0.0236}^{+0.00993+0.0181+0.0244}$&$0.705$&
$0.708_{-0.00940-0.0187- 0.0273}^{+0.00929+0.0187+0.0274}$&$0.705$
\\
$\Omega_m$&$-$&
$0.292_{-0.00993-0.0181-0.0243}^{+0.00987+0.0188+0.0236}$&$0.295$&
$0.292_{-0.00929-0.0187- 0.0274}^{+0.00940+0.0188+0.0273}$&$0.295$
\\
$\sigma_8$&$-$&
$0.805_{-0.0117-0.0242-0.0333}^{+0.0118+0.0232+0.0317}$&$0.800$&
$0.804_{-0.0113-0.0244-0.0332}^{+0.0121+0.0234+0.0323}$&$0.812$
\\
$z_{re}$&$-$&
$10.648_{-1.0777-2.125-2.808}^{+1.0949+2.132+2.697}$&$10.146$&
$10.583_{-1.0354-1.993- 2.735}^{+1.0162+2.0164+2.694}$&$10.362$
\\
$H_0$&$-$&
$68.500_{-0.831-1.632-2.155}^{+0.834+1.581+2.0859}$&$68.474$&
$68.462_{-0.759-1.657-2.385}^{+ 0.887+1.536+2.181}$&$68.479$
\\
${\rm{Age}}/{\rm{Gyr}}$&$-$&
$13.788_{-0.0362-0.0717-0.0997}^{+0.0357+0.0697+0.0923}$&$13.805$&
$13.788_{- 0.0381-0.0705-0.0952}^{+0.0375+0.0737+0.0968}$&$13.791$
\\
\hline\hline
\end{tabular}
\caption{In contrast to the mean values with $1,2,3\sigma$ errors and the best-fit values of the parameters for the $\xi$wCDM model with perturbed $H$ and $\xi$wCDM model with unperturbed $H$ ($Q^{\mu}_A\parallel u^{\mu}_c$), where CMB from Planck + WMAP9, BAO, SNIa, with RSD data sets have been used.}
\label{tab:results-mean-ucdh}
\end{table}
\end{center}
\endgroup

\begingroup
\squeezetable
\begin{center}
\begin{table}
\begin{tabular}{cccccc}
\hline\hline Parameters & Priors & $\xi$wCDM with perturbed $H$ & Best fit & $\xi$wCDM with unperturbed $H$ & Best fit \\ \hline
$\Omega_bh^2$&[0.005,0.1]&
$0.0223_{-0.000248-0.000456-0.000602}^{+0.000244+0.000477+0.000639}$&$0.0225$&
$0.0223_{-0.000245-0.000493-0.000641}^{+0.000265+0.000477+0.000609}$&$0.0224$
\\
$\Omega_ch^2$&[0.01,0.99]&
$0.114_{-0.00183-0.00394-0.00558}^{+0.00209+0.00384+0.00462}$&$0.115$&
$0.114_{-0.00171-0.00393-0.00521}^{+0.00212+0.00366+0.00476}$&$0.115$
\\
$100\theta_{MC}$&[0.5,10]&
$1.0415_{-0.000538-0.00114-0.00146}^{+0.000547+0.00111+0.00143}$&$1.0414$&
$1.0416_{-0.000595-0.00110-0.00138}^{+0.000572+0.00111+0.00143}$&$1.0412$
\\
$\tau$&[0.01,0.8]&
$0.0878_{-0.0136-0.0253-0.0328}^{+0.0127+0.0271+0.0364}$&$0.0981$&
$0.0887_{-0.0134-0.0236-0.0321}^{+0.0119+0.0255+0.0360}$&$0.0830$
\\
$\xi_x$&[0,1]&
$0.00317_{-0.00317-0.00317-0.00317}^{+0.000628+0.00547+0.00929}$&$0.000687$&
$0.00328_{-0.00328-0.00328-0.00328}^{+0.000736+0.00549+0.00816}$&$0.00142$
\\
$w_x$&[-1,0]&
$-0.976_{-0.0239-0.0239-0.0239}^{+0.00480+0.0409+0.0646}$&$-0.996$&
$-0.971_{-0.0292-0.0292-0.0292}^{+0.00644+0.0443+0.0676}$&$-0.994$
\\
$n_s$&[0.5,1.5]&
$0.977_{-0.00541-0.0109-0.0145}^{+0.00540+0.0109+0.0146}$&$0.982$&
$0.977_{-0.00523-0.0106-0.0133}^{+0.00539+0.0108+0.0142}$&$0.978$
\\
${\rm{ln}}(10^{10}A_s)$&[2.4,4]&
$3.0829_{-0.0245-0.0483-0.0626}^{+0.0242+0.0522+0.0678}$&$3.104$&
$3.0851_{-0.0256-0.0471-0.0630}^{+0.0237+0.0494+0.0676}$&$3.0729$
\\
\hline
$\Omega_x$&$-$&
$0.708_{-0.00967-0.0199-0.0263}^{+0.00977+0.0188+0.0240}$&$0.713$&
$0.706_{-0.00953-0.0199-0.0253}^{+0.00961+0.0180+0.0232}$&$0.710$
\\
$\Omega_m$&$-$&
$0.292_{-0.00977-0.0188-0.0238}^{+0.00967+0.0199+0.0264}$&$0.287$&
$0.294_{-0.00961-0.0179-0.0232}^{+0.00953+0.0199+0.0253}$&$0.290$
\\
$\sigma_8$&$-$&
$0.804_{-0.0112-0.0282-0.0948}^{+0.0151+0.0294+0.0336}$&$0.814$&
$0.805_{-0.0122-0.0241-0.0350}^{+0.0123+0.0240+0.0319}$&$0.803$
\\
$z_{re}$&$-$&
$10.706_{-1.0661-2.235-2.959}^{+1.0883+2.217+2.884}$&$11.546$&
$10.793_{-1.0337-2.0605-2.896}^{+1.0497+2.0563+2.869}$&$10.313$
\\
$H_0$&$-$&
$68.503_{-0.773-1.732-2.279}^{+0.901+1.563+2.0218}$&$69.246$&
$68.297_{-0.775-1.745-2.264}^{+0.966+1.571+1.992}$&$68.889$
\\
${\rm{Age}}/{\rm{Gyr}}$&$-$&
$13.787_{-0.0372-0.0668-0.0869}^{+0.0379+0.0698+0.0897}$&$13.760$&
$13.790_{-0.0372-0.0732-0.0927}^{+0.0377+0.0736+0.100}$&$13.784$
\\
\hline\hline
\end{tabular}
\caption{In contrast to the mean values with $1,2,3\sigma$ errors and the best-fit values of the parameters for the $\xi$wCDM model with perturbed $H$ and $\xi$wCDM model with unperturbed $H$ ($Q^{\mu}_A\parallel u^{\mu}_x$), where CMB from Planck + WMAP9, BAO, SNIa, with RSD data sets have been used.}
\label{tab:results-mean-uxdh}
\end{table}
\end{center}
\endgroup

\section{SUMMARY}

In this paper, we considered that the energy transfer rate was proportional to the Hubble parameter and energy density of dark energy, where the expansion rate was the total expansion rate (background plus perturbations), so the background coupling term was $\bar{Q}=3\xi_x\bar{H}\bar{\rho}_x$, and the perturbation part was $\delta Q=3\xi_x\bar{H}\bar{\rho}_x(\delta H/\bar{H}+\delta_x)$. We have deduced the perturbation equations of dark sectors in the rest frame of dark matter or dark energy. The interaction rate was the most characteristic parameter in the coupled model, so we have carried out the analysis on the cosmological implications of this parameter. It was found that the CMB temperature and matter power spectra owned similar evolutions, however, the growth history of structure could break the degeneracy between the $\xi$wCDM model with perturbed $H$ and coupled model with unperturbed $H$.

Then, with CMB from Planck + WMAP9, BAO, SNIa, and RSD measurements, we conducted a full likelihood analysis for the coupled model. The jointing constraint results showed the interaction rate in 3$\sigma$ regions: $\xi_x=0.00305_{-0.00305-0.00305-0.00305}^{+0.000645+0.00511+0.00854}$ for $Q^{\mu}_A\parallel u^{\mu}_c$, and $\xi_x=0.00317_{-0.00317-0.00317-0.00317}^{+0.000628+0.00547+0.00929}$ for $Q^{\mu}_A\parallel u^{\mu}_x$. It meant that the currently available cosmic observations favored small interaction rate which is up to the order of $10^{-3}$, at the same time, the $f\sigma_8(z)$ test could rule out large interaction rate in 1$\sigma$ region. Besides, compared to the previous papers \cite{ref:Yang2014-uc,ref:Yang2014-ux}, we do not find the significant difference on the model parameter space, so it would be very hard even in the future to distinguish the coupled model with perturbed $H$ from coupled model with unperturbed $H$.

\acknowledgements{This work is supported in part by NSFC under the Grants No. 11275035 and "the Fundamental Research Funds for the Central Universities" under the Grants No. DUT13LK01.}

\end{document}